\documentclass{article}

     \PassOptionsToPackage{numbers, compress}{natbib}

\usepackage[preprint]{neurips_2024}

\usepackage[utf8]{inputenc} 
\usepackage[T1]{fontenc}    
\usepackage{hyperref}       
\usepackage{url}            
\usepackage{booktabs}       
\usepackage{amsfonts}       
\usepackage{nicefrac}       
\usepackage{microtype}      
\usepackage{xcolor}         
\usepackage{graphicx}

\title{Quantum Digital Twins for Uncertainty Quantification}

\author{%
  Soronzonbold Otgonbaatar \\
  Remote Sensing Technology Institute\\
  German Aerospace Center (DLR) Oberpfaffenhofen\\
  Weßling 82234 \\
  \texttt{soronzonbold.otgonbaatar@dlr.de} \\
   \And
  Elise Jennings\\
   Applications and Performance Team\\
   ParTec AG\\
   Possartstr. 20, München 81679\\
   \texttt{elise.jennings@par-tec.com} \\
}

\begin{document}

\maketitle

\begin{abstract}
Modern supercomputers can handle resource-intensive computational and data-driven problems in various industries and academic fields. These supercomputers are primarily made up of traditional classical resources comprising CPUs and GPUs. Integrating quantum processing units with supercomputers offers the potential to accelerate and manage computationally intensive subroutines currently handled by CPUs or GPUs. However, the presence of noise in quantum processing units limits their ability to provide a clear quantum advantage over conventional classical resources. Hence, we develop and construct "quantum digital twins," virtual versions of quantum processing units. To demonstrate the potential advantage of quantum digital twins, we create and deploy hybrid quantum ensembles on five quantum digital twins that emulate parallel quantum computers since hybrid quantum ensembles are suitable for distributed computing. Our study demonstrates that quantum digital twins assist in analyzing the actual quantum device noise on real-world use cases and emulate parallel quantum processing units for distributed computational tasks in order to obtain quantum advantage as early as possible. 
\end{abstract}

\section{Quantum Digital Twins}

Today's supercomputers play a crucial role in solving large-scale and data-driven computational problems in fields such as astrophysics \cite{hpc1} and climate science \cite{hpc2}. The fundamental components of supercomputers are parallel central processing units (CPUs) and graphics processing units (GPUs). Human-written software distributes computational tasks to parallel CPUs and GPUs to generate results as quickly as possible. Some computational tasks are even intractable and time-consuming for supercomputers as their precision and the quality of results increase, e.g., high-resolution climate models \cite{hpc2}. In order to overcome this bottleneck, research studies invent and refine quantum algorithms for a few notoriously hard computational problems ranging from simulating the energy of molecules \cite{CheQu}, quantum systems \cite{QA4QD} to machine learning \cite{QA4ML}, making them much faster than traditional classical algorithms. 
Some studies even have proposed the concept of quantum-centric supercomputing for handling complex computational tasks \cite{QPUcentric}, which involves integrating quantum processing units (QPUs) with CPUs and GPUs to harness the best of both traditional and quantum computing resources. Quantum computing approaches can be implemented on either faulty or fault-tolerant QPUs. Currently, available faulty QPUs, also known as noisy-intermediate scale quantum computers \cite{Preskill2018quantumcomputingin}, are primarily designed for processing variational quantum algorithms \cite{Cerezo2021}. However, the main challenge with faulty QPUs is the noise present in quantum bits and gates. Real QPUs for analyzing the practical impact of its noise are even not readily publicly accessible and time-consuming for practically significant problems via cloud services. 
To address this, we propose the development of digital twins of faulty QPUs, known as quantum digital twins, characterized by the noise characteristics of real quantum devices \cite{isakov2021simulationsquantumcircuitsapproximate}. Quantum digital twins can help analyze quantum device data and guide improvements in quantum systems. More importantly, it is accessible 
during real QPU device down time, which may be significant during re-calibration.
The quantum digital twins can be accessed on our GitHub repository \cite{QTwins}.  Toward potential quantum advantage for practical problems, we invent and deploy hybrid quantum ensembles on "N" replica quantum digital twins to emulate parallel QPUs. These can be viewed as deploying "N" hybrid classical-quantum models on "N" parallel QPUs, and the aggregated results are then presented with a certain mean and variance. Hybrid quantum ensembles are the quantum counterparts of conventional ensembles used for uncertainty quantification, making them a promising candidate for parallel computing resources, i.e., quantum-centric supercomputing systems \cite{TEnsemble}. This study may be a crucial step in achieving a quantum advantage for distributed tasks deployed on parallel QPUs at the earliest opportunity.

\begin{figure}[!t]
\centering
\begin{minipage}{.5\textwidth}
  \centering
  \includegraphics[width=\linewidth]{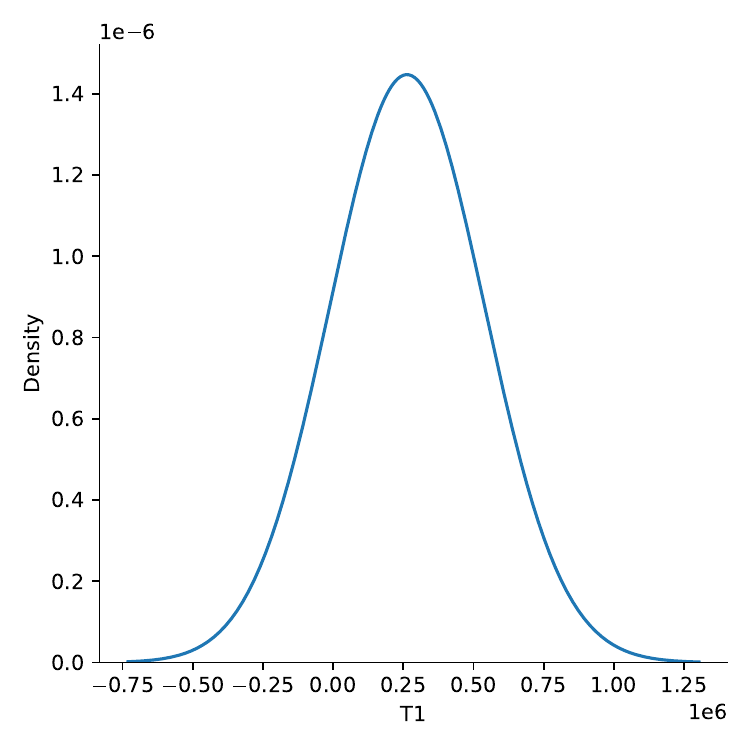}
  \label{fig:test1}
\end{minipage}%
\begin{minipage}{.5\textwidth}
  \centering
  \includegraphics[width=\linewidth]{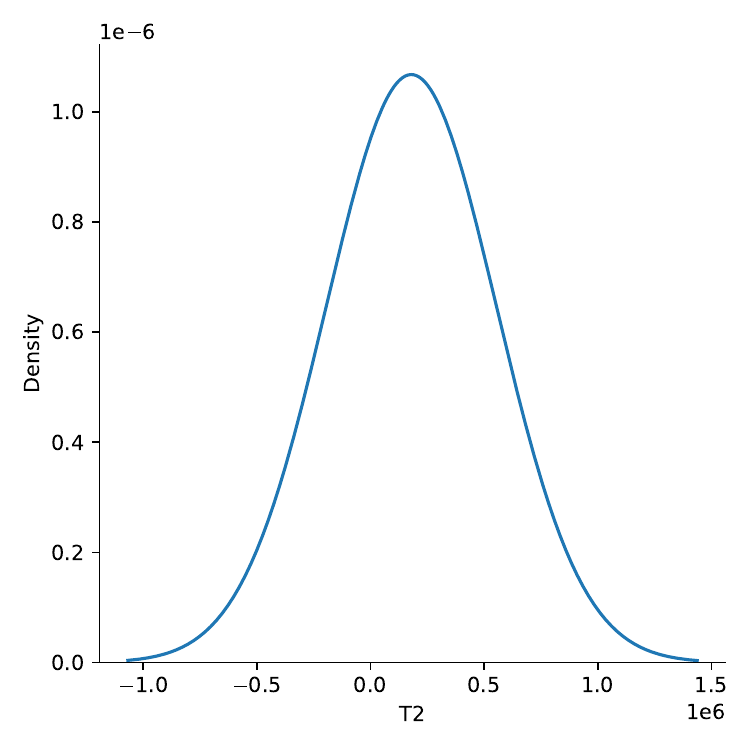}
\end{minipage}
\caption{The distributions of thermal relaxation time $T_1$ and qubit dephasing time $T_2$ of the ``ibm\_sherbrooke'' quantum computer.} \label{fig:test2}
\end{figure}

\section{Uncertainty Quantification}

Deep learning (DL) models are widely used to solve real-world problems in industry and science such as climate modeling \cite{hpc2}. They are preferred for their scalability on large datasets and for their parallelizability on distributed computing resources \cite{Pandey2022, Jared2022}. However, DL models are often seen as uninterpretable black boxes, producing unreliable predictions \cite{Gal2022}. They yield point estimates of predictions with point weights lacking their uncertainty, i.e., lacking explainability due to the uninterpretable black-box paradigm \cite{rudin2018}. There is increasing interest in developing scalable uncertainty quantification techniques for DL models. Hence, the authors of the article \cite{TEnsemble} proposed and invented neural network ensembles more efficient and faster than Bayesian approaches. They can be scaled on large datasets and computed efficiently on GPU tensor cores. This allows DL models to generate predictions with reliable error/uncertainty estimates. Taking advantage of our parallel digital twins of faulty QPUs, we create and embed parameterized quantum circuits in neural network ensembles called hybrid quantum ensembles \cite{Li2024, sozogrsl2021}. We note that our goal is not to demonstrate the superiority of hybrid quantum ensembles over its conventional classical counterpart but to showcase how parallel QPUs can be utilized and benchmarked. Identifying and refining the quantum model that can be distributed to parallel quantum devices with the help of traditional supercomputing systems is crucial to achieving a quantum advantage as early as possible.  

\section{Experiment}

\begin{figure}[!t]  
\centering
 \includegraphics[width=13cm, height=2cm]{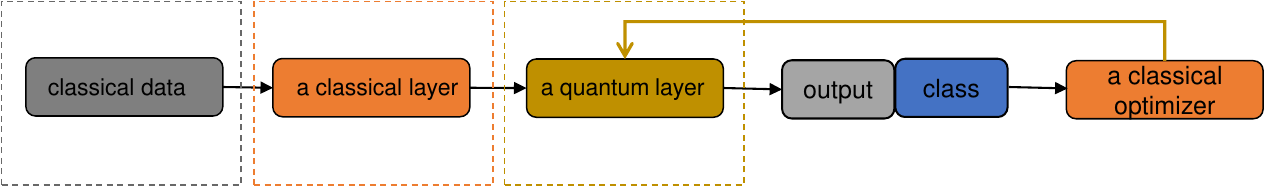}
  \caption{Above is a hybrid classical-quantum model: for deploying hybrid quantum ensembles on ``N'' parallel quantum-centric devices, we replicate a hybrid classical-quantum model ``N'' times, each implemented on a single quantum-centric device separately. The outcome is the average over the ensembles with uncertainty information. The image is taken from the article \cite{sozo2024} (CC-BY-4.0).}\label{fig: fig3}
\end{figure}

\begin{figure}  
\centering
 \includegraphics[width=8cm, height=6cm]{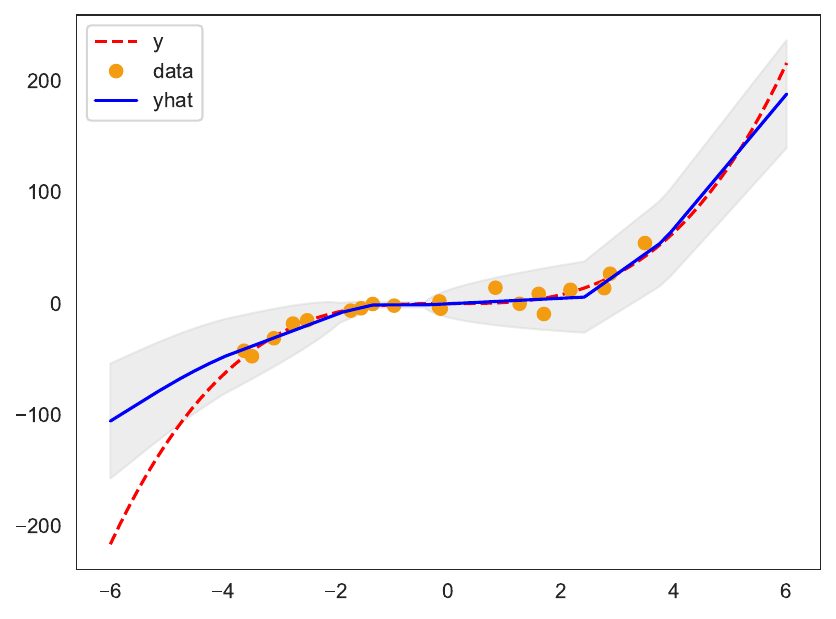}
  \caption{The point estimates of hybrid quantum ensembles with uncertainty information when we deploy the quantum part of hybrid quantum ensembles on five parallel quantum digital twins emulating faulty QPUs. Here, a \textcolor{red}{red} dashed line denotes a true function, a \textcolor{blue}{blue} solid line represents the approximate function with uncertainty quantities shaded by light gray, and \textcolor{orange}{orange} circles represent sampled data points.}\label{fig: fig2}
\end{figure}

We prepare five parallel quantum digital twins based on the actual quantum device data extracted from an ibm\_sherbrooke system with $127$ faulty quantum bits (qubits) \cite{ibm}. The data includes quantum gate duration, quantum gate error, thermal relaxation time $T_1$, qubit dephasing time $T_2$, and qubit readout errors. More importantly, it is versioned on timestamps to avoid the impact of IBM's quantum device calibration. You can access the quantum datasets in the ``QuantumDatabase'' folder on our GitHub repository \cite{QTwins}. The noise of each quantum digital twin is modeled by sampling from quantum data points of the ``ibm\_sherbrooke'' system. We illustrate its distribution of thermal relaxation and qubit dephasing time
in Figure \ref{fig:test2}. Subsequently, we implement hybrid quantum ensembles for uncertainty quantification on five parallel quantum digital twins when using a synthetic regression dataset; hybrid quantum ensembles refer to a collection of hybrid classical-quantum models illustrated in Figure \ref{fig: fig3}. In particular, we execute the quantum layer of our hybrid classical-quantum models on five parallel quantum digital twins simultaneously, whereas deploying the classical layer on CPUs and GPUs. In this setup, a classical layer comprises two hidden layers, each with 100 neurons. In contrast, a quantum layer with three input qubits consists of a feature embedding quantum layer and parameterized single-qubit rotation about the $Y$ axis denoted by $R_Y(\theta)$. Our result demonstrates that hybrid quantum ensembles deployed on parallel digital twins of faulty QPUs in a quantum-centric computing system generate reliable outcomes with uncertainty information. We summarize our main outcome in Figure \ref{fig: fig2}, which illustrates the output of our hybrid quantum ensembles with uncertainty quantities.

\section{Conclusion}

The development of faulty quantum processing units holds promise for providing a quantum advantage over supercomputing systems for certain complex problems. However, such an advantage has yet to be demonstrated for real-world problems. The main challenge lies in the fact that we have small quantum computers, and real-world practical problems require large quantum circuit models. To address this challenge, large quantum circuit models can be cut into a few small quantum circuits  using, e.g., a quantum circuit knitting technique suitable for implementation on parallel small quantum computers \cite{schmitt2024cuttingcircuitsmultipletwoqubit}. Unfortunately, no experimental demonstration has been performed due to the absence of publicly accessible parallel quantum computers. To overcome this challenge,  we construct digital twins of faulty quantum computers containing characteristics derived from real quantum device data. These quantum digital twins enable the emulation of parallel faulty quantum processing units and the deployment of distributed problems on them while also facilitating the analysis of the impact of quantum noise on the problem outcomes. Moving towards practical demonstration, we implement hybrid quantum ensembles for uncertainty quantification using synthetic regression data points on five parallel digital twins of the "ibm\_sherbrooke" quantum computer. These hybrid quantum ensembles generate outcomes with uncertainty information by aggregating several hybrid classical-quantum models. The results of our uncertainty quantification, as shown in Figure \ref{fig: fig2}, demonstrate how hybrid quantum ensembles, operating on five parallel quantum digital twins, can help us examine and compare distributed quantum devices in quantum-centric supercomputing systems. In pursuit of achieving a quantum advantage in addressing real-world problems, we are working on refining the quantum circuit models that represent real-world use cases previously considered intractable by traditional classical techniques. These models can be broken down into smaller quantum circuit models for use in distributed quantum-centric supercomputing systems. More importantly, our quantum device database may help improve the quality of faulty quantum computers using modern artificial intelligence techniques when we aim to tackle challenging scientific problems on quantum computers.    
The quantum digital twins could help us also analyze the actual quantum device noise on real-world computational problems and draw valuable conclusions to guide improvements in quantum systems. 

\bibliographystyle{IEEEtran.bst}
\bibliography{bibliotec}

\end{document}